# Cross relaxation and the bilayer coupling in $Y_2Ba_4Cu_7O_{15}$

H. Monien and T. M. Rice  
Theoretische Physik  
ETH Zürich  
CH-8093 Zürich, Switzerland

There is strong experimental evidence for substantial antiferromagnetic coupling between the $CuO_2$ planes of the bilayers in YBCO compounds. The acoustic mode of the spin excitation spectrum has been measured but the optic mode has not been observed. Theoretical estimates suggest values for the antiferromagnetic interplane coupling as large as 10 meV. It has been proposed that this coupling is responsible for the spin gap in the underdoped YBCO compounds. In the recently synthesized compounds $Y_2Ba_4Cu_7O_{15}$ with alternating single/double chain blocks, it is possible to distinguish between the $CuO_2$ planes in a bilayer. We propose a nuclear magnetic cross relaxation experiment to directly measure the strength of the coupling between the two planes of a bilayer. The temperature dependence and magnitude of this cross relaxation rate is predicted.

Recently the low energy spin dynamics of the high $T_c$ superconductors has attracted a lot of interest. The unusually strong temperature dependence of the measured spin susceptibility, $\chi_s(T)$ of the underdoped compounds rules out a simple Fermi liquid like ground state with a temperature independent spin susceptibility. The decreasing spin susceptibility indicates the proximity to a critical point. Millis and Monien [1] and subsequently Sokol and Pines [2] proposed that this critical point should be identified with the order-disorder transition of the 2D Quantum Heisenberg Antiferromagnet [3] in spite of the fact that doping may well change the universality class of the relevant model. The interesting question arises what drives the system close to this critical point. Millis and Monien concluded from the analysis of NMR data that the spin gap only appears in systems containing $CuO_2$ bilayers. An additional sign that bilayer correlation are important in YBCO comes from the neutron scattering experiments by Tranquada et al. [4] which found nearly complete antiferromagnetic correlations of the spins in the bilayers. The corresponding optic mode has not been observed up to now. Thus one can only obtain a lower limit on the exchange coupling between the planes of the bilayer, $J_\perp > 10$meV, from the neutron scattering experiments. An examination of a model of two antiferromagnetically coupled Heisenberg planes gives that the exchange coupling between the planes, $J_\perp$ at the critcal point is of the order $2.6 \times J$, with $J \sim 1000K$, the exchange coupling in the plane [5,6]. Such an exchange coupling seems much too large for a realistic material [7]. Several papers have appeared to propose mechanisms which could lead to an enhancement of the singlet correlations between the bilayers [8]. Here we propose an experiment to measure $J_\perp$ directly.

In $YBa_2Cu_3O_{6+\delta}$ the two layers form a bilayer separated from the next bilayer by a large distance. Each of the layers of the bilayers is completely equivalent. Recently a new cuprate superconductor has been synthesized in which the $CuO_2$ layers of the bilayers are chemically inequivalent. The unit cell contains two building blocks one with one chain, similar to $YBa_2Cu_3O_7$, the other contains two chains, like $YBa_2Cu_4O_8$. The important point is that the $CuO_2$ layers are attached to a different chemical environment but are otherwise left intact. One partner of the bilayer is attached to the single chain block and the other to the double chain block. The properties of this material have been studied by Stern et al., [9] who find that the resonance frequency of the of the Cu nuclear spins in the two planes of the bilayer are slightly different from each other. This allows them to distinguish the $CuO_2$ planes of the bilayer.

We propose an experiment which utilizes this fact to directly measure the coupling between the planes of the bilayer. This coupling opens a new relaxation channel for the Cu nuclear spins. A nuclear Cu spin, sitting in one particular plane of the bilayers, polarizes the neighboring electronic spins in the same plane which in turn produce an additional field on the Cu nuclear spins in the neighboring plane. Thus this process generates a nuclear spin spin interaction of the form

$$H_{II} = \sum_{ij} I^{(a)}_{i\alpha} V_{a,b}(R_i - R_j) I^{(b)}_{j\beta} \quad (1)$$

The interaction $V_{a,b}(R_i - R_j)$ is defined by its Fourier transform:

$$V_{(a,b)}(q) = F^{(a)}(q) F^{(b)}(q) \chi_s^{(a,b)}(q) \quad (2)$$

where $F^{(a)}(q)$ is the Fourier transform of the hyperfine coupling constant in plane $(a)$, and $\chi^{(a,b)}(q)$ is the static spin susceptibility, $\chi_s^{(ab)}(q) = \int_0^\infty dt\, <S_z^{(a)}(q,t) S_z^{(b)}(q)>$, where $(a)$ and $(b)$ refer to the plane of the bilayer and $q$ is a two dimensional wavevector. If the two nuclear Cu spins are sitting in the same plane this interaction causes the so called transverse relaxation rate $T_2$ [10]. The rate produced by the interplanar nuclear spin spin interaction is given by:

$$\left(\frac{1}{T_2}\right)^2_{\text{cross}} = n_m \sum_q \left[F(q)^2 \chi_s^{(1,2)}(q)\right]^2 \quad (3)$$

where $n_m$ is the density of the active NMR nuclei. This expression agrees with the one derived for the transverse relaxation rate [10] with the important difference that the nuclear spins which are coupled to each other are sitting in different planes of a bilayer.

As long as the two planes of the bilayers are strongly coupled and only the acoustic mode is populated the cross relaxation rate $(1/T_2)_{\text{cross}}$ should be proportional to the nuclear spin - nuclear spin dephasing time $T_2$ which has been measured already. The magnitude of the cross relaxation time $(1/T_2)_{\text{cross}}$ is determined by the same factors as $1/T_2$ with the exception of the spin susceptibility for spins sitting in different planes of the bilayer. A naive estimate gives a reduced value for $\chi^{(1,2)} \sim (J_\perp/J) \chi^{(1,1)}$, where $J$ is the exchange coupling in the planes and $J_\perp$ is the exchange coupling between the the planes of the bilayers. The cross relaxation time would therefore be reduced by a factor of $(J_\perp/J) \sim 0.1$ from the transverse relaxation rate $T_2$. As soon as the temperature is large enough that also the optic mode starts to get populated the two rates should show a strong deviation in the temperature dependence. The cross relaxation time would be strongly reduced and eventually go to zero. The temperature scale at which this should happen is set by the exchange coupling between the planes [1]. This would provide a direct measurement of the exchange coupling between the planes.

The proposed experiment should allow a direct determination of the exchange coupling between the planes of the bilayer and provides an estimate of the energy of the optic mode of the bilayer system.